\documentclass[aps,prl,reprint,superscriptaddress,floatfix,nofootinbib]{revtex4-2}
\usepackage{graphicx}
\graphicspath{{figs/}}
\usepackage{amsmath,amssymb}
\usepackage{bm}
\usepackage{xcolor}

\begin{document}

\title{Interferometric measurement of a quantized rank-2 twist invariant in a frustrated quantum magnet}

\author{Deren~Liu}
\affiliation{Nanyang Technological University, Singapore}
\author{Pinaki~Sengupta}
\affiliation{Nanyang Technological University, Singapore}

\date{\today}

\begin{abstract}
Berry phases can turn global quantum geometry into quantized invariants of quantum many-body systems. However, when the physical low-energy space forms a correlated multiplet, the Berry phase of any single state becomes ill-defined because states can mix within the near-degenerate manifold without shifting the physical subspace. In the two-state multiplet studied here, this basis dependence can drive the single-state phase through zeros, where no phase exists. To resolve this, we treat the multiplet as a whole and label it by the phase of the multiplet twist determinant, the determinant of a compact matrix of twist overlaps between the multiplet states. This phase is basis independent, and with inversion symmetry it equals the Berry phase of the multiplet around the twist loop. Its measurement has so far been obstructed by the depth of controlled state preparation and, in loop-based formulations, by controlled-SWAP cascades. We remove this bottleneck with an architecture in which one unconditional circuit, the loader, prepares the multiplet states and an ancilla qubit controls the twist operator and nothing else. With it we measure the quantized rank-2 twist invariant interferometrically across the frustration-driven dimerization transition of the spin-$1/2$ $J_1$--$J_2$ chain on IBM Heron processors for rings of 6 to 12 sites. Four state preparations then reconstruct the complete projected twist matrix. The processor returns the correct quantized class at all size--coupling points, verified with an independent repetition. A loader trained variationally on Hamiltonian energies alone, with no exact-state information or fidelity in the objective, resolves the complete transition. Quantized multiplet twist invariants of interacting quantum matter at finite size are thus measurable on present-day digital quantum processors.
\end{abstract}

\maketitle

\section{Introduction}

Berry phases become quantized invariants of phases of matter when a symmetry pins them or when they are summed over a closed surface~\cite{berry1984,zak1989,hatsugai2006,resta1998}. When the physically selected low-energy space is a quasi-degenerate multiplet, the phase of any single state can become ill-defined. Within such a near-degenerate pair, no single state can be singled out and followed. Mixing within the multiplet changes the single-state phase and can drive the expectation value of the twist operator defined below through zeros, where no phase exists [Supplemental Material (SM), Sec.~S1]. Individual matrix elements depend on the basis chosen within the multiplet~\cite{wilczek1984}. Such multi-state geometry is central to characterizing symmetry-broken valence-bond solids and, at higher rank, topologically ordered ground spaces~\cite{ntw1985}. In interacting many-body systems, these invariants have been formulated through transported states along closed paths. Examples are Wilson loops of the SU(2) Berry gauge field over twisted boundary conditions~\cite{lee2008} and Berry-curvature forms integrated over multi-parameter families of gapped systems~\cite{kapustin2020}. Each carries states across a family of Hamiltonians, requires the gap to stay open along it, and yields a label of the path rather than of any single Hamiltonian point. Experimentally, quantum geometry has been accessed through single-particle Bloch states~\cite{yu2025,gao2025}, engineered few-level holonomies~\cite{pachos1999,zu2014}, and synthetic atomic gauge fields~\cite{sugawa2021}. To our knowledge it has not been accessed for a correlated many-body multiplet on a digital quantum processor.

On digital quantum processors, Berry-type invariants have been measured for single-particle systems~\cite{niedermeier2024,prr2025,xie2025} and for a periodically driven system~\cite{will2025}, while circuits for many-body invariants of interacting systems have remained proposals or simulations~\cite{sugimoto2023,dominguez2025}. A primary bottleneck is that extracting an off-diagonal matrix element between two correlated states has conventionally required one of two deep constructions: controlled state preparation, in which the deepest part of the circuit runs under the control of the ancilla, or controlled-SWAP permutations between two copies of the qubit register that holds the state~\cite{tan2025}.

Here we measure a multiplet twist invariant of a correlated spin chain on superconducting processors, across the dimerization transition of the frustrated $J_1$--$J_2$ model, with an interferometric architecture that removes this bottleneck. The invariant is defined at a single Hamiltonian point. It is the phase of the determinant of twist overlaps within the low-energy doublet, the two lowest states of the ring. It requires no adiabatic loop or transported family of states, and inversion symmetry pins it to $0$ or $\pi$ at every finite size. The loop-based label requires the multiplet to stay isolated along the entire path. Where it can be computed for comparison, the two labels agree at every size and coupling sampled (SM, Sec.~S9), and the single-point formulation obtains the label from one Hamiltonian point instead of a transported family of states. The measurement removes controlled state preparation, and with it the controlled-SWAP cascades between register copies that set the depth of earlier routes~\cite{tan2025}. The superpositions of the product inputs are prepared before the unconditional loader, the circuit that prepares the two states. The ancilla controls the shallow diagonal twist and nothing else.

\section{Model and quantized invariant}

We apply this architecture to the periodic frustrated antiferromagnetic chain, a ring of $L$ spin-$\tfrac12$ operators $\bm S_q$ with site indices modulo $L$ and frustration ratio $g$,
\begin{equation}
 H=J_1\sum_{q=0}^{L-1}\bm S_q\!\cdot\!\bm S_{q+1}
   +J_2\sum_{q=0}^{L-1}\bm S_q\!\cdot\!\bm S_{q+2},
 \qquad g=J_2/J_1,
 \label{eq:H}
\end{equation}
with $J_1,J_2>0$.  In the thermodynamic limit the chain is critical and gapless for $g<g_c\simeq0.2411$~\cite{okamoto1992,eggert1996} and spontaneously dimerized and gapped for $g>g_c$~\cite{haldane1982}. At the Majumdar--Ghosh point $g=1/2$, the two symmetry-related dimer coverings are exact ground states~\cite{mg1969,mg1969b}.  Its thermodynamics~\cite{leboeuf2003} and ground-state entanglement~\cite{gu2004} have been studied. The wider family hosts multipolar phases for ferromagnetic $J_1$~\cite{parvej2017}, valence-bond-solid transitions in dimerized ferromagnetic chains~\cite{wardyn2023}, and analogous physics for spin-1~\cite{kolezhuk1997}.  The dimerized phase breaks translation symmetry rather than carrying intrinsic long-range-entangled order. Its low-energy doublet nevertheless carries a quantized multiplet label, invariant under basis changes within the doublet (Fig.~\ref{fig:necessity}; SM, Sec.~S1). Inversion symmetry pins this label, as shown below.  The two-state space plays different roles in the two phases. In the dimerized phase it is the ground-state doublet. Below $g_c$ the finite chain has a unique ground state, and the two-state space is the isolated low-energy subspace, gapped from the rest at every size and coupling sampled. The label is then a finite-size spectral diagnostic.

This two-state structure on both sides of the transition is what Lieb--Schultz--Mattis (LSM) theorems demand.  A translation-invariant spin-$\frac12$ ring with
spin-rotation symmetry cannot have a unique gapped ground state in the
thermodynamic limit---at finite $L$ a second eigenstate always lies
within $O(1/L)$ of the ground
state~\cite{lsm1961,affleck1986,hastings2010}---and for sufficiently strong
frustration the $J_1$--$J_2$ chain resolves this alternative through
spontaneous dimerization~\cite{haldane1982}. Consequently the measured doublet can be seen as symmetry-mandated. It is exactly degenerate at the Majumdar--Ghosh point and exponentially near-degenerate on finite rings elsewhere in the dimerized phase~\cite{hastings2010}.

The two lowest orthonormal states in the sector of zero total magnetization, $S^z_{\rm tot}=\sum_qS^z_q=0$, span the selected space. For them we define the twist operator, with the twist angle $\theta=2\pi/L$ per site,
\begin{equation}
 U_\theta=\exp\!\left(i\frac{2\pi}{L}\sum_{q=0}^{L-1}qS_q^z\right).
 \label{eq:Utheta}
\end{equation}
The projected twist matrix of the two lowest states $\vert\psi_a\rangle$ and $\vert\psi_b\rangle$, $a,b\in\{0,1\}$, is
\begin{equation}
 Z_{ab}=\langle\psi_a\vert U_\theta\vert\psi_b\rangle,
 \label{eq:Zab}
\end{equation}
and the rank-2 twist invariant, the phase of the multiplet twist determinant, is
\begin{equation}
 \gamma=\left|\arg\det Z\right|.
 \label{eq:gamma}
\end{equation}
Under a basis change within the selected space by a unitary $2\times2$ matrix $V$, $Z\mapsto V^\dagger ZV$, so $\det Z$ is unchanged.  Quantization follows from bond-centered inversion,
$I: q\mapsto L-1-q$, in three displayed steps.  First, inversion reverses
the twist.  Acting on the exponent,
\begin{align}
 I\Big(\sum_{q}q\,S^z_q\Big)I^\dagger
   &=\sum_{q}q\,S^z_{L-1-q}
    =\sum_{q'}(L-1-q')\,S^z_{q'}\nonumber\\
   &=(L-1)\,S^z_{\rm tot}-\sum_{q'}q'\,S^z_{q'},
 \label{eq:invexponent}
\end{align}
so in the $S^z_{\rm tot}=0$ sector
\begin{equation}
 I\,U_\theta\,I^\dagger=U_\theta^\dagger .
 \label{eq:invtwist}
\end{equation}
Second, if the selected space is inversion invariant, inversion acts inside it by a unitary $B$ with $I|\psi_b\rangle=\sum_c B_{cb}|\psi_c\rangle$. Inserting Eq.~\eqref{eq:invtwist} into each matrix element gives
\begin{align}
 Z_{ab}&=\langle\psi_a|I^\dagger\,U_\theta^\dagger\,I|\psi_b\rangle
   =\sum_{cd}B^*_{ca}B_{db}\,\langle\psi_c|U_\theta^\dagger|\psi_d\rangle
   \nonumber\\
  &=\big(B^\dagger Z^\dagger B\big)_{ab}.
 \label{eq:invZ}
\end{align}
Third, taking the determinant of Eq.~\eqref{eq:invZ} and using
$|\det B|=1$,
\begin{equation}
 \det Z=\det\!\big(B^\dagger Z^\dagger B\big)=\det Z^\dagger=(\det Z)^*,
 \label{eq:pinning}
\end{equation}
pinning every nonzero determinant to $\gamma=0$ or $\pi$.  This proves the quantization measured here (SM, Sec.~S1). The conditions under which a multiplet is selected and the dependence on the origin of the twist are developed separately for arbitrary rank.  For every finite size and
coupling sampled, exact diagonalization gives $\gamma=0$ at $g=0,0.1,0.2$ and $\gamma=\pi$ at $g=0.3,0.4,0.5$. The hardware tests this quantized classification and brackets the transition, while the thermodynamic $g_c$ itself is known independently~\cite{okamoto1992}.

 Consider a state $|\phi(\alpha)\rangle = \cos\alpha|\psi_0\rangle + \sin\alpha|\psi_1\rangle$ in the exact $L=6$ low-energy plane, the two-dimensional space spanned by $|\psi_0\rangle$ and $|\psi_1\rangle$. We compare two quantities derived from the twist matrix $Z$: the single-state expectation value $z_\phi(\alpha) = \langle\phi(\alpha)|U_\theta|\phi(\alpha)\rangle$, which varies with $\alpha$, and $\det Z$, which does not.

Because the diagonal elements vanish ($Z_{00}=Z_{11}=0$) and the off-diagonal elements are symmetric ($Z_{10}=Z_{01}$), the single-state expectation value reduces to $z_\phi(\alpha) = \sin(2\alpha)\,Z_{01}$. It depends on the state chosen within the plane, vanishes at the eigenstates themselves ($\alpha=0$ and $\pi/2$), and its phase jumps by $\pi$ across each zero. At $g=0.2$ the matrix is $Z=-0.765793\,i\sigma_x$, with $\sigma_x$ the Pauli matrix in the basis $\{|\psi_0\rangle,|\psi_1\rangle\}$, and the single-state phase is $\pm\pi/2$ between the zeros. At $g=0.4$ the matrix is $Z=+0.660169\,\sigma_x$ and the single-state phase is $0$ on some intervals of $\alpha$ and $\pi$ on others. 

The determinant is unchanged, because a basis change within the plane acts as $Z\mapsto V^\dagger ZV$. It is $\det Z=+0.586439$ at $g=0.2$ and $-0.435824$ at $g=0.4$ for all $\alpha$, invariant to within $10^{-15}$ under random unitary rotations of the basis [Fig.~\ref{fig:necessity}; SM, Sec.~S1].

\begin{figure}[t]
\centering
\includegraphics[width=\columnwidth]{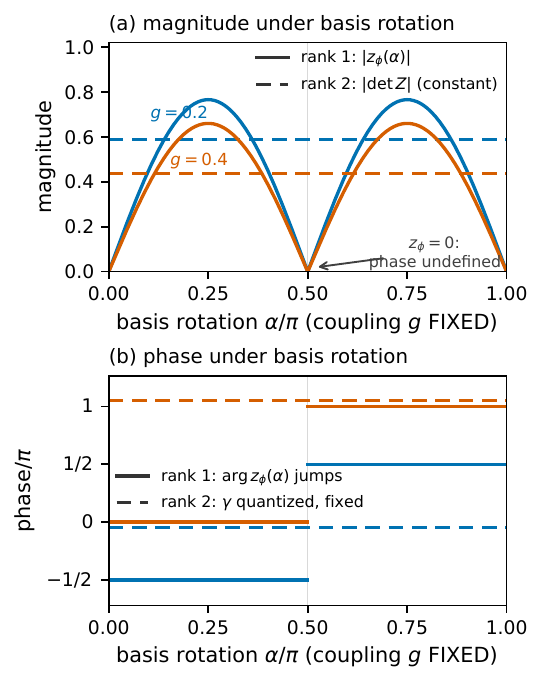}
\caption{\textbf{Rank 1 is basis dependent; rank 2 is not.}
Exact $L=6$ single-state twist expectation $z_\phi=\langle\phi(\alpha)|U_\theta|\phi(\alpha)\rangle$ under basis rotations $\alpha$ inside the selected low-energy plane at fixed coupling, one curve from each phase ($g=0.2$ and $0.4$). The dependence on $g$ itself, including the jump at the transition, is Fig.~\ref{fig:protocol}(b).  Solid: rank 1; dashed: rank 2.
(a) The rank-1 magnitude $|z_\phi|$ collapses to exact zero at $\alpha=0,\pi/2,\pi$, where the single-state phase is undefined. The rank-2 determinant magnitude is unchanged. (b) The rank-1 phase is locked to one of the two values allowed by inversion between zeros and jumps by $\pi$ across each zero, so no single state carries a basis-independent phase. The rank-2 determinant phase $\gamma$ stays at its quantized value (dashed lines,
offset slightly from $0$ and $\pi$ for visibility), invariant to within $10^{-15}$ (SM, Sec.~S1).}
\label{fig:necessity}
\end{figure}

\section{Interferometric measurement architecture}

\begin{figure*}[t]
\centering
\includegraphics[width=\textwidth]{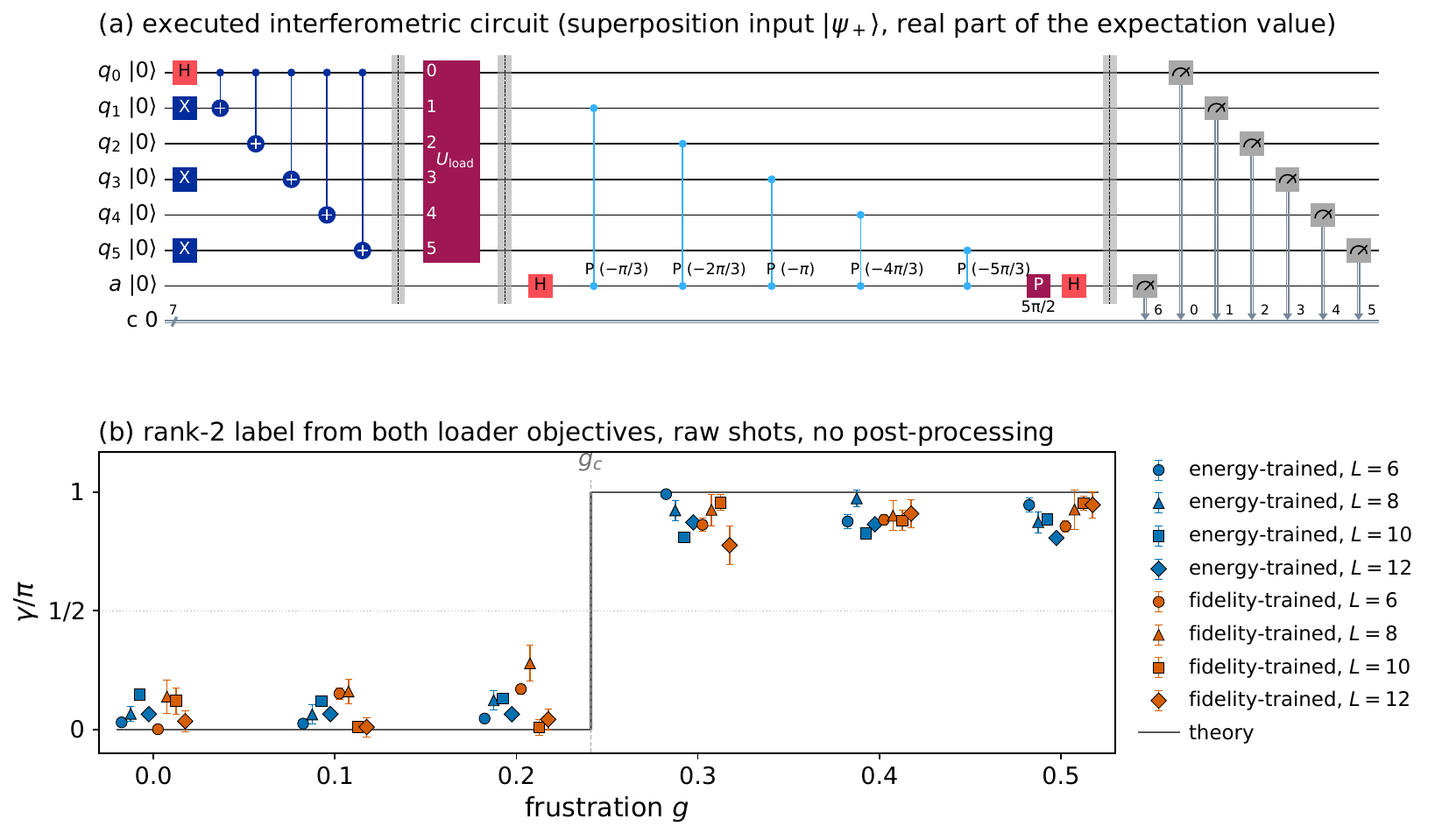}
\caption{\textbf{Hamiltonian-trained interferometric measurement.}
(a) The executed interferometric circuit at $L=6$ (the setting measuring
the real part of the expectation value, superposition input
$|\psi_+\rangle$ shown).  The branch superposition is
prepared on product inputs by one Hadamard and a CNOT fan-out before the loader (the two branches of qubit $q_0$ carry the two N\'eel input strings). The correlated loader $U_{\rm load}$ is unconditional. The ancilla controls only the product of controlled-phase gates of Eq.~\eqref{eq:twistfactor}, with the $q=0$ gate of angle zero omitted, and is read out after the closing Hadamard. The imaginary part is measured by inserting one $S^\dagger$ on the ancilla before that Hadamard. System measurements verify the conserved $S^z_{\rm tot}=0$ sector. (b) Two independent preparations, Hamiltonian-energy training and exact-state fidelity training, at $L=6$ (circles), $L=8$ (triangles), $L=10$ (squares), and $L=12$ (diamonds), from the raw shots without symmetry filtering. All resolve the complete $0\!\to\!\pi$ change.  Error bars are
one standard deviation from 2,000 multinomial bootstrap replicates.  The
gray step is the theoretical classification, jumping at the known
thermodynamic $g_c$ (dashed line). Points are offset horizontally for visibility.}
\label{fig:protocol}
\end{figure*}

 One unconditional loader $U_{\text{load}}$ maps two orthogonal N\'eel input strings $|r_0\rangle=|1010\cdots\rangle$ and $|r_1\rangle=|0101\cdots\rangle$ onto an orthonormal pair $\{\vert \psi_0\rangle, \vert \psi_1\rangle\}$ trained to span the two-state low-energy space. By linearity, the same loader applied to a superposition of the two inputs prepares the same superposition of $\vert\psi_0\rangle$ and $\vert\psi_1\rangle$, in particular the balanced superpositions
\begin{equation}
\vert \psi _{+}\rangle =\frac{\vert \psi _{0}\rangle +\vert \psi _{1}\rangle }{\sqrt{2}},\qquad \vert \psi _{+i}\rangle =\frac{\vert \psi _{0}\rangle +i\vert \psi _{1}\rangle }{\sqrt{2}}.
\end{equation}
For each state $\vert \chi_k\rangle \in \{\vert \psi_0\rangle,\vert \psi_1\rangle,\vert \psi_+\rangle,\vert \psi_{+i}\rangle\}$, the ancilla qubit acts as a one-qubit interferometer. Following a Hadamard gate on the ancilla, the controlled twist acts on the path where the ancilla is $1$, yielding the joint state
\begin{equation}
\vert \Phi_k\rangle=\frac{\vert 0\rangle\otimes\vert \chi_k\rangle +\vert 1\rangle\otimes U_\theta\vert \chi_k\rangle}{\sqrt2}.
\label{eq:twopath}
\end{equation}
Tracing out the system register, the $L$ qubits that hold the state, leaves the ancilla in a reduced state whose coherence is the overlap of the two paths,
\begin{equation}
\rho_{\text{a}}={\text{Tr}}_{\text{sys}}\,\vert \Phi_k\rangle\langle\Phi_k\vert =\frac12\begin{pmatrix} 1 & \langle\chi_k\vert U_\theta^\dagger\vert \chi_k\rangle\\ \langle\chi_k\vert U_\theta\vert \chi_k\rangle & 1 \end{pmatrix}, \label{eq:ancillarho} \end{equation} and the Pauli expectations of the ancilla read these coherences out, $\langle X_{\text{a}}\rangle=2\,{\text{Re}}\,(\rho_{\text{a}})_{10}$ and $\langle Y_{\text{a}}\rangle=2\,{\text{Im}}\,(\rho_{\text{a}})_{10}$. The closing Hadamard alone gives the real part, and one extra $S^{\dagger}$ before it gives the imaginary part. Two circuits per preparation therefore give the complex expectation value
\begin{equation}
d_k=\langle\chi_k\vert U_\theta\vert \chi_k\rangle =\langle X_{\text{a}}\rangle+i\langle Y_{\text{a}}\rangle.
\label{eq:hadamard}
\end{equation}
The deep loader is applied once and is not conditioned on the ancilla. Only $U_\theta$ is controlled [Fig.~\ref{fig:protocol}(a)]. With $S^z_q=1/2-n_q$, where $n_q$ is the computational bit of site $q$, the controlled twist factorizes into phase gates,
\begin{equation}
C(U_\theta)=P_{\text{a}}\!\left(\frac{\pi}{L}\sum_q q\right) \prod_{q=0}^{L-1}CP_{{\text{a}},q}\!\left(-\frac{2\pi q}{L}\right),
\label{eq:twistfactor}
\end{equation}
where $C(U_\theta)$ is the twist controlled by the ancilla, $P_{\text{a}}(\varphi)$ a phase gate on the ancilla that supplies the constant spin-$1/2$ phase factor, and $CP_{{\text{a}},q}(\varphi)$ a controlled-phase gate of angle $\varphi$ between the ancilla and site $q$.  Since the $q=0$ gate has angle zero, the only controlled element of the circuit is a string of exactly $L-1$ two-qubit gates.

The four expectation values determine $Z$ through a polarization identity. Expanding the two superposition expectations entry by entry,
\begin{align}
d_+&=\tfrac12\big(Z_{00}+Z_{11}\big)+\tfrac12\big(Z_{01}+Z_{10}\big), \nonumber\\
d_{+i}&=\tfrac12\big(Z_{00}+Z_{11}\big)+\tfrac{i}{2}\big(Z_{01}-Z_{10}\big),
\label{eq:quadexpand}
\end{align}
while $d_0=Z_{00}$ and $d_1=Z_{11}$ directly. Solving for the off-diagonal elements gives
\begin{align}
Z_{01}+Z_{10}&=2d_+-(d_0+d_1),\nonumber\\
Z_{01}-Z_{10}&=i(d_0+d_1)-2id_{+i}, \label{eq:polarization} \end{align} so same-state expectation values at superposed arguments determine the two-state matrix elements. The processor supplies every complex entry of $Z$ by interference, and the classical final step is two linear equations and a $2\times2$ determinant. The construction extends to any rank $r$: the $r$ basis states plus two superpositions per unordered pair give $r+2\binom{r}{2}=r^2$ coherent preparations and, with real and imaginary circuits, $2r^2$ circuits for the complete projected matrix. This is a constructive rule rather than an optimality bound. With the same common loader from known orthogonal inputs, direct branch interference is an equivalent single-register realization of the same architecture (SM, Secs.~S4--S5). In it the ancilla itself selects which input string enters the loader.

\section{Hamiltonian-trained state preparation}

The loader is a brick wall of magnetization-conserving exchange blocks. Each block acts on one bond $(i,j)$ with two angles $s$ and $c$ and is built from the two-qubit rotations $R_{xx}$, $R_{yy}$, and $R_{zz}$ about the Pauli operators $X_i$, $Y_i$, and $Z_i$,
\begin{equation}
 G(s,c)=R_{xx}(s)R_{yy}(s)R_{zz}(2c)
 \label{eq:gate}
\end{equation}
The blocks act on even, odd, and periodic-wrap bonds, followed by $R_z$ rotations.  The equal $XX$ and $YY$ angles enforce $S^z$ conservation. The factor 2 makes $c$ the unit coefficient of $Z_iZ_j$ in the exponent,
$G=\exp[-\tfrac{is}{2}(X_iX_j{+}Y_iY_j)-icZ_iZ_j]$ (SM, Sec.~S2, Fig.~2).
The $ZZ$ generator takes it beyond the free-fermion matchgate manifold, and one
unitary prepares the complete orthogonal plane rather than fitting two
unrelated states.  For $r$ repetitions on $L$ sites it has $3Lr$ parameters
(executed schedules: SM, Sec.~S2).

At $L=6$ the loader, with parameter vector $\bm\beta$ (90 parameters), is trained by minimizing
\begin{equation}
 C(\bm\beta)=
 \sum_{c=0}^{1}\langle r_c|U_{\rm load}^\dagger H U_{\rm load}|r_c\rangle,
 \label{eq:energycost}
\end{equation}
a unitary subspace-search objective that preserves input orthogonality by
construction~\cite{nakanishi2019}.  This objective targets the required subspace, and the bound is short.  Let the exact eigenstates $|E_i\rangle$ of $H$ in the sector have energies $E_0\le E_1\le E_2\le\cdots$, and let $\lambda_i= \sum_{c=0}^{1}|\langle E_i|U_{\rm load}|r_c\rangle|^2$ be the occupation of $|E_i\rangle$ by the prepared plane, so that
\begin{equation}
 C=\sum_i\lambda_iE_i,\qquad
 \sum_i\lambda_i=2,\qquad 0\le\lambda_i\le1,
 \label{eq:kyfanoccupations}
\end{equation}
and let $\ell=\sum_{i\ge2}\lambda_i$ be the total weight outside the true
low-energy plane.  Writing $\lambda_0=1-u$ and $\lambda_1=1-v$ with
$u,v\ge0$ and $u+v=\ell$,
\begin{align}
 C-(E_0+E_1)&=\sum_{i\ge2}\lambda_iE_i-uE_0-vE_1\nonumber\\
 &\ge \ell E_2-uE_0-vE_1
 \;\ge\;(E_2-E_1)\,\ell,
 \label{eq:kyfanbound}
\end{align}
using $E_2-E_0\ge E_2-E_1$ in the last step.  Hence $C\ge E_0+E_1$, and
whenever $E_1<E_2$ the energy excess
$\varepsilon=C-(E_0+E_1)$ bounds the leakage by
$\ell\le\varepsilon/(E_2-E_1)$: minimizing energy alone provably targets
the invariant-carrying subspace (SM, Sec.~S2).  Exact eigenvectors, their fidelities, and the target invariant are absent from the objective, initialization, stopping rule, and parameter selection. Multiple deterministic starts are ranked by energy alone, and the lowest-energy parameters are frozen before any hardware shot.  The frozen $L=6$ loaders lie within $10^{-6}$ of the exact two-state energy sum at every coupling. Two diagnostics computed after the parameters were frozen certify that the invariant-carrying plane was hit: the leakage of $H$ out of the prepared plane and the subspace fidelity of the prepared plane to the exact one (SM, Sec.~S2).  The same protocol certifies the $L=8$, $10$, and $12$ loaders at all six couplings with subspace fidelities above $0.96$. All three sets classify correctly on hardware at every coupling (Fig.~\ref{fig:protocol}; SM, Sec.~S2).

The training was then repeated against a different objective. Loaders compiled against exact eigenvectors provide the $L=6,8,10,12$ runs with the same ansatz, and their ideal fidelities establish finite-depth representability of the complete doublet through $L=12$ (SM, Sec.~S2).

\section{Hardware results}

All circuits ran on two IBM Heron processors, \texttt{ibm\_marrakesh} and \texttt{ibm\_kingston}, with standard error suppression. Figure~\ref{fig:protocol} gives the central result: all 48 measured classifications across the 24 size--coupling points are correct. The energy-trained and the fidelity-compiled runs at every size, spread over the two processors, resolve the transition between $g=0.2$ and $0.3$.

This agrees with finite-size theory. On an $L$-site ring the label jumps at the crossing of the lowest triplet with the second singlet, $g_x(L)$~\cite{okamoto1992,eggert1996}. Exact diagonalization puts $g_x(L)$ between $0.242$ and $0.250$ for $L=6$ to $20$, converging to the thermodynamic $g_c$ as $1/L^2$ (SM, Sec.~S1). The couplings $g=0.2$ and $0.3$ therefore bracket $g_x(L)$ at every size measured.

\begin{figure}[t] \centering \includegraphics[width=\columnwidth]{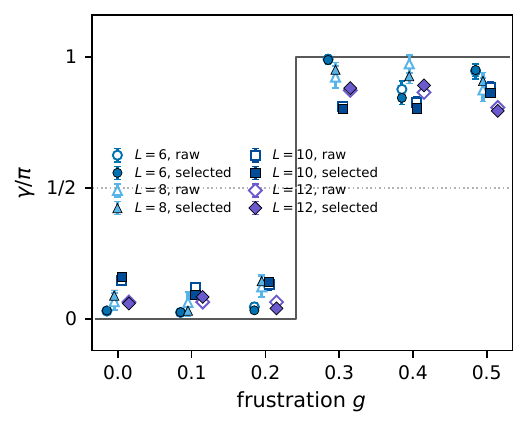} \caption{\textbf{Symmetry filtering is not needed for the classification.} The energy-trained hardware runs assign the correct phase at every coupling from the raw shots alone (open symbols) as well as after $S^z_{\rm tot}=0$ selection (filled symbols). No point changes side of the $\gamma/\pi = 1/2$ decision line (dotted) between the raw and the filtered analysis, so no classification depends on the symmetry filter. The fidelity-compiled runs behave the same way (SM, Sec.~S7). Points are offset horizontally for visibility.}
\label{fig:raw}
\end{figure}

Figure~\ref{fig:protocol}(b) uses the raw shots with no symmetry filter. With $S^z_{\rm tot}=0$ symmetry verification~\cite{bonet2018} applied instead, every run again classifies 6/6. All 48 classifications hold with and without postselection (Fig.~\ref{fig:raw}; SM, Sec.~S7).

\begin{figure}[t] \centering \includegraphics[width=\columnwidth]{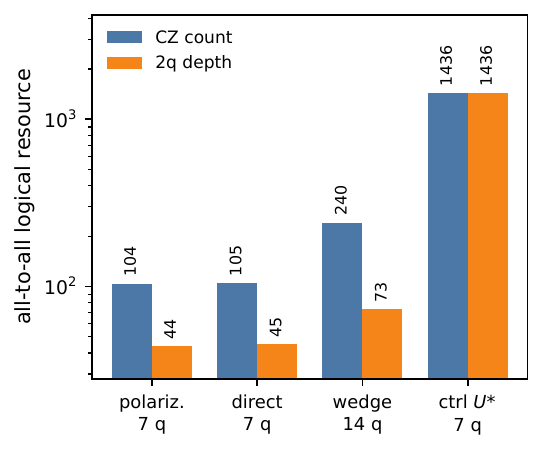} \caption{\textbf{Architectural saving.} All-to-all logical resources for the same frozen energy-trained loader ($L=6$, $g=0.3$). The executed polarization circuit (``polariz.'') and its direct-branch equivalent (``direct'') have nearly identical gate counts and depths. The two-copy determinant construction (``wedge'', the antisymmetric two-copy state) doubles the register. A loader placed entirely under ancilla control (``ctrl $U^*$'', a stress test rather than an optimized alternative) is an order of magnitude deeper.}
\label{fig:resources}
\end{figure}

The saving that makes this feasible is architectural. Compiled on an equal logical footing with the same frozen loader, the executed polarization circuit costs 104 CZ gates on 7 qubits, and its direct-branch counterpart costs nearly the same. The two-copy determinant construction, which reads $\det Z$ as one expectation value on two copies of the register, doubles the register to 14 qubits. A loader placed entirely under ancilla control, a stress test rather than an optimized alternative, is an order of magnitude deeper [Fig.~\ref{fig:resources}; SM, Sec.~S5]. Device routing leaves the executed circuits well within the capability of the processors used (SM, Sec.~S5). The architecture uses one unconditional loader and one system register, and places only the shallow diagonal twist under control.

\section{Discussion}

To our knowledge, this is the first interferometric measurement on a digital quantum processor of a quantized twist invariant defined over a correlated many-body multiplet. Three results compose the paper, and each is independent of the others.

The first is algorithmic. Off-diagonal matrix elements between correlated states have required placing the deep state preparation under ancilla control, or controlled-SWAP cascades between register copies. Here the branch superpositions are prepared on product states by a shallow fan-out before the loader, the deep loader runs unconditionally, and the only controlled operation is a shallow product of controlled-phase gates that implements the diagonal twist. The processor computes every complex matrix element by interference: each measured real or imaginary part is a relative phase between the two ancilla paths, and the off-diagonal elements follow from the polarization identity. The construction extends to any rank at $2r^2$ circuits, with two equivalent single-register wirings (SM, Secs.~S4--S5).

The second is physical: the full rank of the two-state low-energy space is necessary for a basis-independent phase and modulus. The twist phase of any single state inside the near-degenerate pair is basis-dependent, undefined at the zeros of its expectation value, and driven through jumps by basis rotations. The determinant of the multiplet is basis-invariant. Pinned by bond-centered inversion to $0$ or $\pi$, the rank-2 invariant separates the two sides of the transition on finite rings, and the measured determinant does so correctly at every size and coupling sampled.

The third is variational. The loader can be trained on Hamiltonian energies alone. The energy excess above the exact two-state energy sum bounds the leakage out of the target plane. The training is demonstrated at every size studied, $L=6$ to $12$, and the frozen circuits classify correctly on hardware at every coupling. No exact eigenvector or target phase entered training or parameter selection at any stage.

The evidence is deliberately redundant. Two types of training, one on Hamiltonian energies alone and one compiled against exact eigenstates, were spread over two processors and return the same classification at every shared point. The raw counts alone classify correctly before any symmetry filtering.

The measured object is the determinant sector of the multiplet geometry at a single Hamiltonian point, not a path-ordered holonomy. No loop of twisted Hamiltonians is prepared or transported, which keeps the measurement shallow where loop-based constructions require additional depth~\cite{tan2025}. The energy objective was evaluated by exact simulation here, but it is measurable on the device in three global bases, so on-device training is limited by hardware time. No barren-plateau immunity or asymptotic optimizer-convergence claim is made~\cite{cerezo2021}. Quantized multiplet twist invariants of interacting quantum matter at finite size are thus measurable on present-day digital quantum hardware.

\emph{Data and code availability.}  Raw counts, job identifiers, frozen
analysis code, parameter sets, and the exact submitted circuits are
preserved in the project archive and will be available on reasonable request.

\begin{acknowledgments}
We acknowledge the use of IBM Quantum services for this work. The authors used an AI assistant to improve the presentation and for language editing. All suggestions were reviewed and approved by the authors. We acknowledge the support of the Singapore Ministry of Education Academic Research Fund's Tier 3 Grant MOE-MOET32023-0003 (PS) and Tier 2 Grant MOET2EP50123-0021 (DL).
\end{acknowledgments}


\begin{thebibliography}{99}
\bibitem{berry1984} M.~V.~Berry, Proc.~R.~Soc.~Lond.~A \textbf{392}, 45 (1984).
\bibitem{zak1989} J.~Zak, Phys.~Rev.~Lett.~\textbf{62}, 2747 (1989).
\bibitem{hatsugai2006} Y.~Hatsugai, J.~Phys.~Soc.~Jpn.~\textbf{75}, 123601 (2006).
\bibitem{resta1998} R.~Resta, Phys.~Rev.~Lett.~\textbf{80}, 1800 (1998).
\bibitem{wilczek1984} F.~Wilczek and A.~Zee, Phys.~Rev.~Lett.~\textbf{52}, 2111 (1984).
\bibitem{ntw1985} Q.~Niu, D.~J.~Thouless, and Y.-S.~Wu, Phys.~Rev.~B \textbf{31}, 3372 (1985).

\bibitem{lee2008} S.-S.~Lee and S.~Ryu, Phys.~Rev.~Lett.~\textbf{100}, 186807 (2008).
\bibitem{kapustin2020} A.~Kapustin and L.~Spodyneiko, Phys.~Rev.~B \textbf{101}, 235130 (2020).
\bibitem{yu2025} J.~Yu, B.~A.~Bernevig, R.~Queiroz, E.~Rossi, P.~T\"orm\"a, and B.-J.~Yang, npj Quantum Mater.~\textbf{10}, 101 (2025).
\bibitem{gao2025} A.~Gao, N.~Nagaosa, N.~Ni, and S.-Y.~Xu, Rev.~Mod.~Phys.~\textbf{98}, 035005 (2026).
\bibitem{pachos1999} J.~Pachos, P.~Zanardi, and M.~Rasetti, Phys.~Rev.~A \textbf{61}, 010305(R) (1999).
\bibitem{zu2014} C.~Zu, W.-B.~Wang, L.~He, W.-G.~Zhang, C.-Y.~Dai, F.~Wang, and L.-M.~Duan, Nature \textbf{514}, 72 (2014).
\bibitem{sugawa2021} S.~Sugawa, F.~Salces-Carcoba, Y.~Yue, A.~Putra, and
I.~B.~Spielman, npj Quantum Inf.~\textbf{7}, 144 (2021).
\bibitem{niedermeier2024} M.~Niedermeier, M.~Nairn, C.~Flindt, and
J.~L.~Lado, Phys.~Rev.~Research \textbf{6}, 043288 (2024).
\bibitem{prr2025} H.-C.~Chang, H.-C.~Hsu, and Y.-C.~Lin,
Phys.~Rev.~Research \textbf{7}, 013043 (2025).
\bibitem{xie2025} Q.~Xie, K.~Seki, T.~Shirakawa, and S.~Yunoki,
arXiv:2504.07499 (2025).
\bibitem{will2025} M.~Will \textit{et al.}, Nature \textbf{645}, 348 (2025).
\bibitem{sugimoto2023} T.~Sugimoto, arXiv:2304.13408 (2023).
\bibitem{dominguez2025} S.~Dom\'inguez-Calder\'on, M.~Niedermeier,
J.~L.~Lado, and P.~M.~Vecsei, Phys.~Rev.~B \textbf{114}, 065139 (2026).
\bibitem{tan2025} K.~C.~Tan, D.~Liu, K.~Bharti, D.~Bhowmick, and
P.~Sengupta, Phys.~Rev.~B \textbf{112}, 014108 (2025).
\bibitem{okamoto1992} K.~Okamoto and K.~Nomura, Phys.~Lett.~A \textbf{169}, 433 (1992).
\bibitem{eggert1996} S.~Eggert, Phys.~Rev.~B \textbf{54}, R9612 (1996).
\bibitem{haldane1982} F.~D.~M.~Haldane, Phys.~Rev.~B \textbf{25}, 4925 (1982); \textbf{26}, 5257(E) (1982).
\bibitem{mg1969} C.~K.~Majumdar and D.~K.~Ghosh,
J.~Math.~Phys.~\textbf{10}, 1388 (1969).
\bibitem{mg1969b} C.~K.~Majumdar and D.~K.~Ghosh, J.~Math.~Phys.~\textbf{10}, 1399 (1969).
\bibitem{leboeuf2003} P.~A.~Leboeuf, M.~E.~Gouv\^ea, and A.~S.~T.~Pires, J.~Magn.~Magn.~Mater.~\textbf{264}, 175 (2003).
\bibitem{gu2004} S.-J.~Gu, H.~Li, Y.-Q.~Li, and H.-Q.~Lin, Phys.~Rev.~A \textbf{70}, 052302 (2004).
\bibitem{parvej2017} A.~Parvej and M.~Kumar, Phys.~Rev.~B \textbf{96}, 054413 (2017).
\bibitem{wardyn2023} J.~Wardyn, S.~Nishimoto, and C.~E.~Agrapidis, Phys.~Rev.~B \textbf{108}, 205111 (2023).
\bibitem{kolezhuk1997} A.~Kolezhuk, R.~Roth, and U.~Schollw\"ock, Phys.~Rev.~B \textbf{55}, 8928 (1997).
\bibitem{lsm1961} E.~Lieb, T.~Schultz, and D.~Mattis, Ann.~Phys.~(N.Y.)~\textbf{16}, 407 (1961).
\bibitem{affleck1986} I.~Affleck and E.~H.~Lieb, Lett.~Math.~Phys.~\textbf{12}, 57 (1986).
\bibitem{hastings2010} M.~B.~Hastings, arXiv:1008.5137; published in \emph{Quantum Theory from Small to Large Scales}, Les Houches Session XCV (Oxford University Press, Oxford, 2012).
\bibitem{nakanishi2019} K.~M.~Nakanishi, K.~Mitarai, and K.~Fujii,
Phys.~Rev.~Research \textbf{1}, 033062 (2019).
\bibitem{bonet2018} X.~Bonet-Monroig, R.~Sagastizabal, M.~Singh, and
T.~E.~O'Brien, Phys.~Rev.~A \textbf{98}, 062339 (2018).
\bibitem{cerezo2021} M.~Cerezo, A.~Sone, T.~Volkoff, L.~Cincio, and
P.~J.~Coles, Nat.~Commun.~\textbf{12}, 1791 (2021).
\end{thebibliography}
\end{document}

\title{Supplemental Material: Interferometric measurement of a quantized rank-2 twist invariant in a frustrated quantum magnet}

\author{Deren~Liu}
\affiliation{Nanyang Technological University, Singapore}
\author{Pinaki~Sengupta}
\affiliation{Nanyang Technological University, Singapore}
\date{\today}
\maketitle

\section{S1. Twist convention and rank-2 invariant}
The ring of the main text has $L$ sites,
\begin{equation}
 H=J_1\sum_{q=0}^{L-1}\bm S_q\!\cdot\!\bm S_{q+1}+J_2\sum_{q=0}^{L-1}\bm S_q\!\cdot\!\bm S_{q+2},
 \qquad g=J_2/J_1,
 \label{eq:S_H}
\end{equation}
with $\bm S_q$ the spin-$\tfrac12$ operators, site indices modulo $L$, and $g_c\simeq0.2411$ the transition coupling of the thermodynamic limit. We encode the computational bit $n_q\in\{0,1\}$ of site $q$ through $S_q^z=1/2-n_q$ and write $S^z_{\rm tot}=\sum_qS^z_q$ for the total magnetization. The twist operator is
\begin{equation}
 U_\theta=\exp\!\left(i\frac{2\pi}{L}\sum_{q=0}^{L-1}qS_q^z\right).
 \label{eq:S_twist}
\end{equation}
Its eigenvalue on a measured system bitstring $z$ is
\begin{equation}
 u(z)=\exp\!\left[i\frac{\pi}{L}\sum_q q
                   -i\frac{2\pi}{L}\sum_q qn_q(z)\right].
 \label{eq:S_eigenvalue}
\end{equation}
For an orthonormal basis $\{|\psi_a\rangle\}_{a=0,1}$ of the selected low-energy space, the span of the two lowest eigenstates of $H$ in the $S^z_{\rm tot}=0$ sector,
\begin{equation}
 Z_{ab}=\langle\psi_a|U_\theta|\psi_b\rangle,
 \qquad \gamma=|\arg\det Z|.
\end{equation}
A basis change gives $Z\mapsto V^\dagger ZV$ and leaves $\det Z$ invariant.
The measured scalar is the determinant, or $U(1)$, sector of the rank-2 Berry
geometry.  The experiment does not reconstruct a path-ordered noncommuting
holonomy.  For the finite chains used here, the exact reference is
$\gamma/\pi=(0,0,0,1,1,1)$ at $g=(0,0.1,0.2,0.3,0.4,0.5)$.  The known
thermodynamic $g_c\simeq0.2411$ is not fitted from the hardware data.

The zero-or-$\pi$ quantization used in the experiment follows directly from
bond-centered inversion.  Let $I$ map site $q$ to $L-1-q$.  Then
\begin{align}
 I^\dagger\left(\sum_q qS_q^z\right)I
 &=(L-1)S_{\rm tot}^z-\sum_q qS_q^z,\nonumber\\
 I^\dagger U_\theta I&=U_\theta^\dagger
 \qquad (S_{\rm tot}^z=0).
 \label{eq:S_inversion}
\end{align}
If the selected low-energy space is inversion invariant and $B$ represents
$I$ within it, Eq.~\eqref{eq:S_inversion} gives
$B^\dagger ZB=Z^\dagger$.  Taking determinants yields
$\det Z=(\det Z)^*$. Every nonzero determinant is therefore real and has phase zero or $\pi$.  A separate rank-$N$ analysis develops the more
general selection, translation, and twist-origin conditions.

In the momentum basis the pinning is explicit and it identifies the label. Let $|\psi_0\rangle$ and $|\psi_1\rangle$ be the members of the doublet with momenta $0$ and $\pi$, $T|\psi_0\rangle=|\psi_0\rangle$ and $T|\psi_1\rangle=-|\psi_1\rangle$, where $T$ is the translation that moves every site by one. Translation shifts the exponent of the twist, $T\big(\sum_qqS^z_q\big)T^\dagger=\sum_q(q-1)S^z_q+LS^z_0$, so in the $S^z_{\rm tot}=0$ sector and for spin $\tfrac12$
\begin{align}
 TU_\theta T^\dagger=U_\theta\,e^{2\pi iS^z_0}=-U_\theta .
 \label{eq:S_TUT}
\end{align}
The twist shifts the momentum by $\pi$, and the diagonal entries vanish, $Z_{00}=Z_{11}=0$. Bond-centered inversion reverses the translation, $ITI^\dagger=T^\dagger$, so it keeps the momenta $0$ and $\pi$ and acts on the two members by their parities, $I|\psi_0\rangle=\eta_0|\psi_0\rangle$ and $I|\psi_1\rangle=\eta_\pi|\psi_1\rangle$ with $\eta_0,\eta_\pi=\pm1$, so that $B={\rm diag}(\eta_0,\eta_\pi)$ represents $I$ within the doublet. Inserting Eq.~\eqref{eq:S_inversion} into the off-diagonal entry gives
\begin{align}
 Z_{01}&=\langle\psi_0|IU_\theta^\dagger I^\dagger|\psi_1\rangle=\eta_0\eta_\pi\langle\psi_0|U_\theta^\dagger|\psi_1\rangle=\eta_0\eta_\pi Z_{10}^{*},\nonumber\\
 \det Z&=-Z_{01}Z_{10}=-\eta_0\eta_\pi|Z_{10}|^{2}.
 \label{eq:S_parity}
\end{align}
The label is the product of the two inversion parities, $\gamma=0$ when they differ and $\gamma=\pi$ when they agree, and $|\det Z|=|Z_{10}|^{2}$ is the squared modulus of the one entry that connects the two momenta. At $g\le0.2$ the doublet is the singlet ground state and the $S^z_{\rm tot}=0$ member of the lowest triplet, which have opposite parities, and at $g\ge0.3$ it is two singlets of equal parity. The jump of the label is therefore the singlet--triplet level crossing by which the transition is located in level spectroscopy (Okamoto and Nomura 1992, cited in the main text), seen here as a parity change of one member of the doublet. The determinant is defined without reference to inversion, is measured as one number with a phase and a modulus, and extends to multiplets of any rank.

  With $\sigma_x$ the Pauli matrix in the basis $\{|\psi_0\rangle,|\psi_1\rangle\}$, the exact $L=6$ projected matrices are, to the shown precision,
\begin{align}
 Z(g=0.2)&=-0.765793i\,\sigma_x,& \det Z&=+0.586439,\\
 Z(g=0.4)&=+0.660169\,\sigma_x,& \det Z&=-0.435824.
\end{align}
For $|\phi(\alpha)\rangle=\cos\alpha|\psi_0\rangle+\sin\alpha|\psi_1\rangle$ at $g=0.2$, the phase of $\langle\phi|U_\theta|\phi\rangle$ is $\pm\pi/2$ and is undefined at its zeros.  At $g=0.4$, it switches between
zero and $\pi$ through zeros.  The rank-2 phases remain zero and $\pi$ under
arbitrary $U(2)$ rotations of the selected basis.  In 500 random rotations,
the largest numerical determinant changes were $1.12\times10^{-15}$ and
$6.69\times10^{-16}$, respectively.

\begin{figure}[t]
\centering
\includegraphics[width=0.55\textwidth]{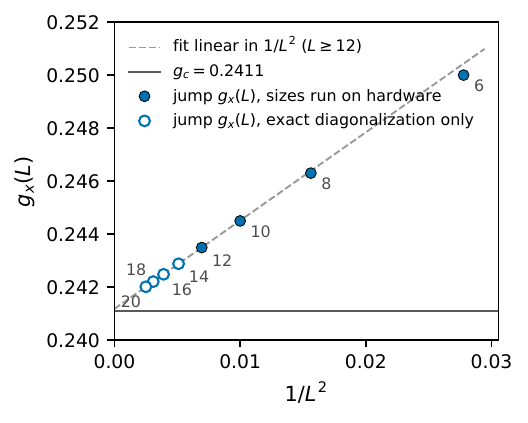}
\caption{Finite-size position of the jump.  Coupling $g_x(L)$ at which the
rank-2 label of the $L$-site ring jumps from $0$ to $\pi$, from exact
diagonalization (the crossing of the lowest triplet with the second
singlet), against $1/L^2$.  Filled symbols: sizes run on hardware, each
sampled at $g=0.2$ and $0.3$, one coupling on either side of its $g_x(L)$.
The dashed line is a fit linear in $1/L^2$ to $L\geq12$, with intercept $0.24118$. The solid line is the accepted thermodynamic value $g_c=0.2411$ (Okamoto and Nomura 1992, Eggert 1996, cited in the main text).}
\label{fig:S_cross}
\end{figure}

On a finite ring the label jumps from $0$ to $\pi$ at the coupling $g_x(L)$
where the lowest triplet crosses the second singlet, so that the two lowest
levels change from a singlet--triplet pair to the dimer doublet.
Table~\ref{tab:S_cross} lists $g_x(L)$ from exact diagonalization in the $S^z_{\rm tot}=0$ sector (levels identified by spin-flip parity, bisection to $10^{-6}$, Fig.~\ref{fig:S_cross}).  A fit linear in $1/L^2$ to $L\geq12$ has residuals below
$2\times10^{-6}$ and intercept $0.24118$, against the accepted
thermodynamic value $0.2411$ cited in the main text.

\begin{table}[h]
\caption{Finite-size jump position $g_x(L)$ of the rank-2 label.}
\label{tab:S_cross}
\begin{ruledtabular}
\begin{tabular}{ccccccccc}
$L$ & 6 & 8 & 10 & 12 & 14 & 16 & 18 & 20\\
\hline
$g_x(L)$ & 0.25000 & 0.24630 & 0.24449 & 0.24349 & 0.24288 & 0.24248 & 0.24220 & 0.24201\\
\end{tabular}
\end{ruledtabular}
\end{table}

\section{S2. Hamiltonian-only and fidelity-trained loaders}
Let $|r_0\rangle=|1010\cdots\rangle$ and $|r_1\rangle=|0101\cdots\rangle$ be the two N\'eel input strings.  The loader $U_{\rm load}$ is a single unitary acting on both inputs, so its outputs remain orthogonal.  Each repetition applies the block
\begin{equation}
 G(s,c)=R_{xx}(s)R_{yy}(s)R_{zz}(2c)
 =\exp\!\left[-\frac{is}{2}(X_iX_j+Y_iY_j)-icZ_iZ_j\right]
\end{equation}
on each bond $(i,j)$ of the even, odd, and periodic-wrap sets, with two angles $s$ and $c$ per bond, followed by one $R_z$ layer.  The block is built from the two-qubit rotations $R_{xx}$, $R_{yy}$, and $R_{zz}$ about the Pauli operators $X_i$, $Y_i$, and $Z_i$.  The $L=6$ loader uses five repetitions, 30 interaction blocks, 90 parameters,
and 90 logical CZ gates under the exact three-CZ decomposition of each block.
In general the ansatz has $3Lr$ parameters for $r$ repetitions.  On an ideal
ring, each repetition has a constant number of parallel bond stages and hence
logical depth proportional to $r$. Routing the periodic bond on a heavy-hex device adds architecture-dependent overhead.  The executed schedule is $r=(5,5,5,8)$ at $L=(6,8,10,12)$ for the
fidelity-compiled runs and $r=(5,5,6,8)$ at $L=(6,8,10,12)$ for the
energy-trained loaders.  These finite values are reported directly
and are not fitted to an asymptotic scaling law.

\begin{figure*}[t]
\centering
\includegraphics[width=\textwidth]{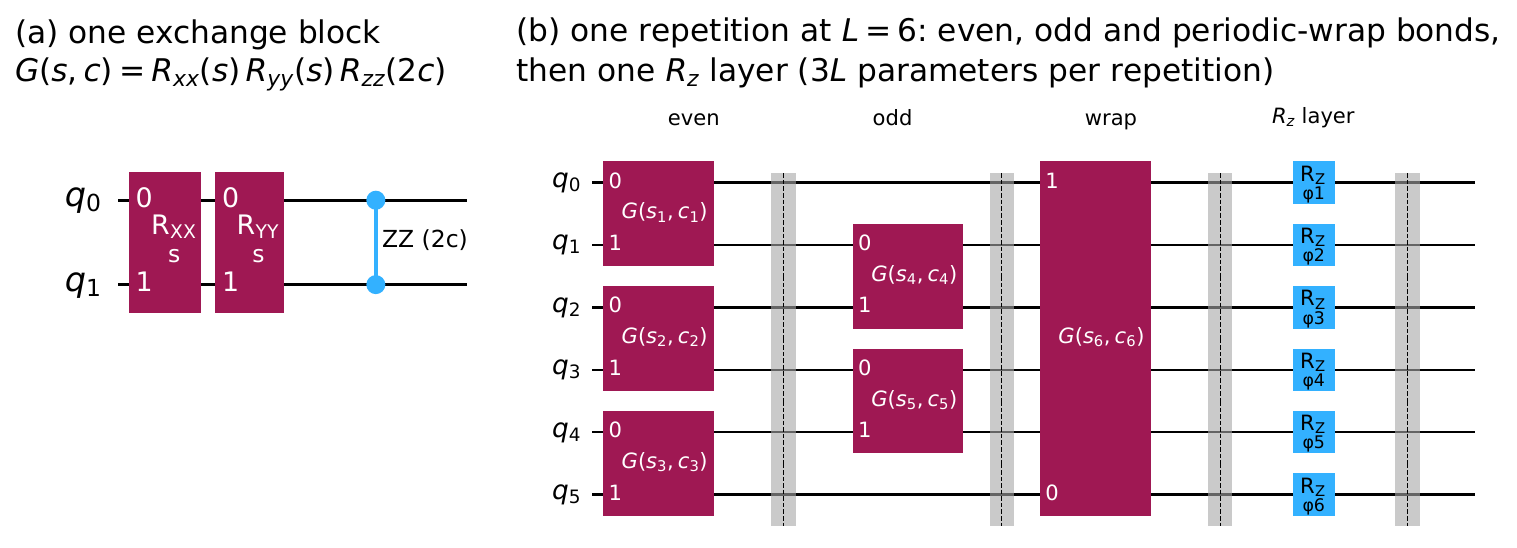}
\caption{\textbf{The loader ansatz.}  (a) One exchange block $G(s,c)=R_{xx}(s)R_{yy}(s)R_{zz}(2c)$. The equal $XX$ and $YY$ angles enforce $S^z$ conservation (on parallel spins $X_iX_j+Y_iY_j$ annihilates, on antiparallel spins it generates the flip-flop). The factor 2 makes $c$ the unit coefficient of $Z_iZ_j$ in the exponent.
(b) One repetition at $L=6$: blocks with independent parameters
$(s_k,c_k)$ on the even bonds, the odd bonds, and the periodic-wrap bond
$(5,0)$, followed by one layer of $R_z(\varphi_q)$ rotations. Each repetition has $3L$ parameters, $3Lr$ in total.  The executed loaders stack
$r=5$ such repetitions at $L=6$ ($r=8$ at $L=12$).}
\label{fig:S_ansatz}
\end{figure*}

For Hamiltonian-only training with parameter vector $\bm\beta$, the cost is
\begin{equation}
 C(\bm\beta)=\sum_{c=0}^{1}\langle r_c|U_{\rm load}^\dagger HU_{\rm load}|r_c\rangle.
 \label{eq:S_cost}
\end{equation}
Minimizing this sum targets the lowest two-dimensional space without selecting a basis inside it. With $E_0\le E_1\le E_2\le\cdots$ the sector energies and $E_2-E_1$ the isolation gap, the energy excess above $E_0+E_1$ bounds the weight outside that space [main text, Eq.~(19)].  This is a statement about
the objective, not a guarantee that a finite-depth ansatz or local optimizer
reaches its optimum.

At every coupling, five
deterministic starts were run.  For later couplings, two were warm-started
from the preceding Hamiltonian-only solution and three were independently
random.  Selection used the smallest value of Eq.~\eqref{eq:S_cost} and nothing else.
Exact eigenvectors, exact phases, subspace fidelities, and variances were not
used in the objective, initialization, stopping rule, or selection.

The energy was evaluated by exact classical statevector simulation rather than
by hardware sampling.  This does not require the exact eigenstates.  The same
objective is measurable on a quantum processor using three global bases for each output:
the $X$ basis gives all $XX$ bond terms, $Y$ gives all $YY$ terms, and $Z$
gives all $ZZ$ terms.  Thus one objective evaluation requires six circuit
settings.

Table~\ref{tab:S_training} gives diagnostics evaluated only after parameter
selection.  The subspace fidelity is
$F_{\rm sub}=\tfrac12\operatorname{Tr}(P P_{\rm exact})$, where $P$ projects onto the two prepared outputs and $P_{\rm exact}$ onto the two exact lowest eigenstates.  Because Eq.~\eqref{eq:S_cost} may return an
arbitrary rotated basis inside the correct plane, individual output variances
are not basis-invariant certificates.  We instead report
\begin{equation}
 \epsilon_H=\lVert(\mathbf 1-P)HP\rVert_F,
 \label{eq:S_Hleak}
\end{equation}
the Frobenius norm of the part of $H$ that maps the prepared plane outside itself, which vanishes exactly when the prepared plane is invariant under $H$.

\begin{table*}[t]
\caption{Hamiltonian-only $L=6$ training.  $\Delta E=C-(E_0+E_1)$, and $F_{\rm old}$ is the post-hoc subspace fidelity to the previously executed
fidelity-trained loader.  None of the last four columns entered training or
parameter selection.}
\label{tab:S_training}
\begin{ruledtabular}
\begin{tabular}{cccccc}
$g$ & $\Delta E$ & $\epsilon_H$ & $F_{\rm sub}$ & $F_{\rm old}$ & ideal $\gamma/\pi$\\
\hline
0.0 & $1.11\times10^{-8}$ & $1.74\times10^{-4}$ & 0.999999997 & 0.999999996 & 0.00000034\\
0.1 & $4.93\times10^{-10}$& $3.32\times10^{-5}$ & 1.000000000 & 0.999999729 & 0.00000514\\
0.2 & $1.79\times10^{-10}$& $1.94\times10^{-5}$ & 1.000000000 & 0.999999995 & 0.00000254\\
0.3 & $5.58\times10^{-7}$ & $7.99\times10^{-4}$ & 0.999999049 & 0.999999034 & 0.99917545\\
0.4 & $3.12\times10^{-11}$& $7.17\times10^{-6}$ & 1.000000000 & 0.999999994 & 0.99999926\\
0.5 & $8.86\times10^{-11}$& $1.12\times10^{-5}$ & 1.000000000 & 0.999999997 & 0.99999922\\
\end{tabular}
\end{ruledtabular}
\end{table*}

At $g=0.3$, continuing the $g=0.2$ parameters alone converged to a higher-energy branch with $C=-4.28815$, while independent starts found $C=-4.40000$. Parameter continuation does not follow the low-energy plane through a level reordering. This is why the protocol combines multistart training with energy-only selection and the residual of Eq.~\eqref{eq:S_Hleak}.

The same protocol extends to $L=8$ (120 parameters, five repetitions,
the same optimizer and energy-only selection rule, with larger multistart sweeps). The loaders were trained by exact simulation and then executed on \texttt{ibm\_marrakesh}, as described below. At five of the six couplings the multistart protocol converges directly. At the near-transition point $g=0.3$ the continuation from $g=0.2$ converges to the $(E_0,E_2)$ plane, with energy excess $0.095$ close to the isolation gap $E_2-E_1=0.089$ and subspace fidelity $0.499$. Continuing instead from the dimerized side ($g=0.5\to0.4\to0.3$) resolves it.  Selecting the lowest
energy per coupling gives, for $g=(0.0,\dots,0.5)$: energy excess $(2.8, 1.9,
2.7, 15.7, 6.5, 0.25)\times10^{-3}$, subspace fidelity $(0.99950, 0.99967,
0.99944, 0.99604, 0.99815, 0.99992)$, and ideal determinant phases
classifying 6/6 correctly. The leakage bound is weakest where the isolation gap is small, as expected.  The frozen $L=8$ set was subsequently
executed on \texttt{ibm\_marrakesh} (48 circuits at 8{,}000 shots, XY4 dynamical decoupling, measurement twirling, routed two-qubit counts 214--239, the transpiled circuits archived) and classifies 6/6 both from raw counts and after $S^z_{\rm tot}=0$ postselection. The raw-count phases are the $L=8$ points of main-text Fig.~2(b) (postselected values: Sec.~S7).

The same protocol extends to $L=10$ with six repetitions (180 parameters;
continuation in both directions, lowest energy per coupling).  At the
near-transition point $g=0.3$, five repetitions leave a competing plane with
the same translation quantum numbers but determinant phase $0$ as the energy
minimum (isolation gap $E_2-E_1=0.074$). Six repetitions make the correct plane the minimum, and energy-only selection then certifies all six couplings.  For
$g=(0.0,\dots,0.5)$ the frozen set has energy excess $(6.8, 5.6, 7.8,
68.1, 37.7, 8.4)\times10^{-3}$, subspace fidelity $(0.9986, 0.9988,
0.9980, 0.9654, 0.9864, 0.9965)$, excess over isolation gap $(0.02, 0.03,
0.13, 0.92, 0.18, 0.02)$, and ideal determinant phases classifying 6/6
correctly.  It was executed on \texttt{ibm\_kingston} (48 circuits at 8{,}000 shots, XY4 dynamical decoupling, measurement twirling, routed two-qubit counts 273--318, the transpiled circuits archived) and classifies 6/6 both from raw counts and after $S^z_{\rm tot}=0$ postselection. The raw-count phases are the $L=10$ points of main-text Fig.~2(b) (postselected values: Sec.~S7).

The same protocol extends to $L=12$ with eight repetitions (288 parameters;
continuation in both directions, lowest energy per coupling).  For $g=(0.0,\dots,0.5)$ the frozen set has energy excess $(7.0,
3.8, 8.1, 35.4, 22.9, 2.1)\times10^{-3}$, subspace fidelity $(0.9985,
0.9992, 0.9982, 0.9836, 0.9915, 0.9992)$, excess over isolation gap $(0.03,
0.03, 0.17, 0.56, 0.12, 0.01)$, and ideal determinant phases classifying
6/6 correctly. At $g=0.3$ the lowest-energy solution lies $0.028$ below the exact energy of the competing plane, so eight repetitions suffice at this size.  It was executed on \texttt{ibm\_kingston} (48 circuits at 32{,}000 shots, XY4 dynamical decoupling, measurement twirling, routed two-qubit counts 319--388, the transpiled circuits archived) and classifies 6/6 both from raw counts and after $S^z_{\rm tot}=0$ postselection. The raw-count phases are the $L=12$ points of main-text
Fig.~2(b) (postselected values: Sec.~S7).

The simulator, its gradients, the Hamiltonian, the compiled gates, and the
ideal controlled-twist circuits were cross-checked against independent constructions to within numerical rounding error.

The fidelity-compiled runs use loaders fitted jointly to the two exact eigenvectors, with five repetitions for $L=6,8,10$ and eight for $L=12$.
Because these parameters use the exact states, they test circuit scaling, not state preparation from the Hamiltonian alone. The Hamiltonian-only runs show
that exact-state fitting is unnecessary for preparing the physical subspace
or obtaining the classification.

For these loaders, a single shared unitary is applied to two orthogonal inputs at each of the 24 size--coupling points. The 48 squared overlaps of its two outputs with the corresponding exact eigenstates have mean 0.987669 and minimum 0.938393.  These statevector quantities establish finite-depth
representability at the tested points. They are not hardware state fidelities, and they do not show that energy-only training would find the
same parameters.

At the near-transition point $L=10$, $g=0.3$, the two same-depth values are 0.961966 and 0.965565. At $L=12$, $g=0.3$, they are 0.978074 and 0.979071.

Trainability and expressibility are kept separate.  The local Hamiltonian
cost, the fixed magnetization sector, small-angle initialization, and
continuation between nearby couplings are chosen to retain physically relevant
gradients.  The $L=6$ multistart result demonstrates convergence for the
tested scan, including at the coupling where plain continuation failed.  Because the symmetry-sector dimension and
executed circuit depth still grow with $L$, no barren-plateau-free or
asymptotic optimizer-convergence claim is made.

Without $R_{zz}$ the ansatz is a number-conserving matchgate-like family.
In the tested setting ($L=6$, $g=0.5$) the joint fit of the doublet then reaches one state exactly and the other not at all, with fidelities $(0.00,\,1.00)$ unchanged from 5 to 20 repetitions. With $R_{zz}$ both states are reached to $1.000$ at five repetitions. This is why the $ZZ$ generator is included.  This observation is specific to the ansatz, targets,
optimizer, and depths tested.

\section{S3. Executed controlled-twist circuit}
The four required inputs $|r_0\rangle$, $|r_1\rangle$, $(|r_0\rangle+|r_1\rangle)/\sqrt2$, and $(|r_0\rangle+i|r_1\rangle)/\sqrt2$ give, by linearity and in that order, the four loaded states $|\chi_k\rangle$ of the main text.  The two superposed inputs are prepared
before the loader by a Hadamard, an optional phase, and a CNOT fan-out.  No
second system register is used.

After loading $|\chi_k\rangle$, an ancilla is prepared by $H|0\rangle$.  The
controlled twist is
\begin{equation}
 C(U_\theta)=P_{\rm a}\!\left(\frac{\pi}{L}\sum_q q\right)
 \prod_{q=0}^{L-1}CP_{{\rm a},q}\!\left(-\frac{2\pi q}{L}\right),
 \label{eq:S_controlled}
\end{equation}
where $C(U_\theta)$ is the twist controlled by the ancilla, $P_{\rm a}(\varphi)$ a phase gate on the ancilla that supplies the constant spin-$1/2$ phase, and $CP_{{\rm a},q}(\varphi)$ a controlled-phase gate of angle $\varphi$ between the ancilla and site $q$.  The real circuit
applies a final ancilla Hadamard.  The imaginary circuit applies $S^\dagger$
and then the final Hadamard.  With the recorded convention,
\begin{equation}
 d_k=\langle Z_{\rm a}\rangle_{\rm real}
       +i\langle Z_{\rm a}\rangle_{\rm imag}
     =\langle\chi_k|U_\theta|\chi_k\rangle.
\end{equation}
Every circuit measures the ancilla and all $L$ system qubits.  The controlled twist is the only controlled many-body operation. $U_{\rm load}$ is not conditioned on the ancilla.

For a measured bit string $x$ with count $c(x)$, the leftmost bit $x_{\rm a}$ is the ancilla.  The raw expectation value estimator is
\begin{equation}
 \widehat z=\frac{\sum_xc(x)(-1)^{x_{\rm a}}}{\sum_xc(x)}.
\end{equation}
The symmetry-selected estimator used in Sec.~S7 and in main-text Fig.~3
restricts both sums to system strings with exactly $L/2$ ones, the conserved
$S^z_{\rm tot}=0$ sector.

\section{S4. Polarization identity}
For $d_0=Z_{00}$, $d_1=Z_{11}$,
\begin{align}
 d_+&=\tfrac12(Z_{00}+Z_{11}+Z_{01}+Z_{10}),\\
 d_{+i}&=\tfrac12(Z_{00}+Z_{11}+iZ_{01}-iZ_{10}).
\end{align}
Hence
\begin{align}
 s_{01}&=2d_+-(d_0+d_1),&
 a_{01}&=i(d_0+d_1)-2id_{+i},\\
 Z_{01}&=\tfrac12(s_{01}+a_{01}),&
 Z_{10}&=\tfrac12(s_{01}-a_{01}),\\
 \widehat\gamma&=\left|-\arg(d_0d_1-Z_{01}Z_{10})\right|.
 \label{eq:S_estimator}
\end{align}
One coupling therefore requires four system preparations, each read in its real-part and imaginary-part circuit: eight circuits.  No exact vector or exact phase enters
Eq.~\eqref{eq:S_estimator} after the loader parameters are frozen.

For rank $r$, measure the $r$ basis-state expectations. For each unordered pair $i<j$, measure also the expectations of $(|\psi_i\rangle+|\psi_j\rangle)/\sqrt2$ and $(|\psi_i\rangle+i|\psi_j\rangle)/\sqrt2$.  The corresponding pair of
polarization equations reconstructs both ordered entries $Z_{ij}$ and
$Z_{ji}$.  There are
\begin{equation}
 r+2{r\choose2}=r^2
\end{equation}
coherent preparations and $2r^2$ real- and imaginary-part circuits per
parameter point.  Random complex matrices at ranks 2 through 6 were
reconstructed with maximum entrywise error below $3.4\times10^{-16}$.  This
is a constructive scaling rule, not an optimality theorem. Direct branch interference can be equivalent when the same common loader from known
orthogonal inputs is available.

\section{S5. Circuit identities and resource comparison}
We compared four constructions using the same frozen Hamiltonian-only $L=6$, $g=0.3$, five-repetition loader: the executed polarization circuit, its direct-branch equivalent, the two-copy determinant construction (the antisymmetric two-copy state, called the wedge below), and a loader placed entirely under ancilla control.  Circuits were transpiled to the gate set \{CZ, $\sqrt X$, $R_z$, $X$\} with all-to-all connectivity at optimization level 3.  This isolates logical architecture from device routing.

\begin{table*}[t]
\caption{Logical circuit comparison for the same loader.  The controlled
loader is a direct compilation stress test, not an optimized alternative or a
lower bound.  Counts are worst-case over the required real- and
imaginary-part circuits.}
\label{tab:S_resources}
\begin{ruledtabular}
\begin{tabular}{lrrrr}
construction & active qubits & CZ gates & two-qubit depth & total depth\\
\hline
present polarization & 7 & 104 & 44 & 169\\
equivalent direct branch & 7 & 105 & 45 & 170\\
two-copy determinant construction (wedge) & 14 & 240 & 73 & 270\\
controlled complete loader, stress test & 7 & 1436 & 1436 & 5522\\
\end{tabular}
\end{ruledtabular}
\end{table*}

The direct-branch construction prepares the input branch superposition before
the same unconditional loader.  Its matrix elements agree with the direct
projected matrix to $1.44\times10^{-15}$, explaining its near equality with
polarization in Table~\ref{tab:S_resources}.  Thus polarization is a
convenient reconstruction, not a unique depth saving over every
single-register alternative.

\begin{figure*}[t]
\centering
\includegraphics[width=\textwidth]{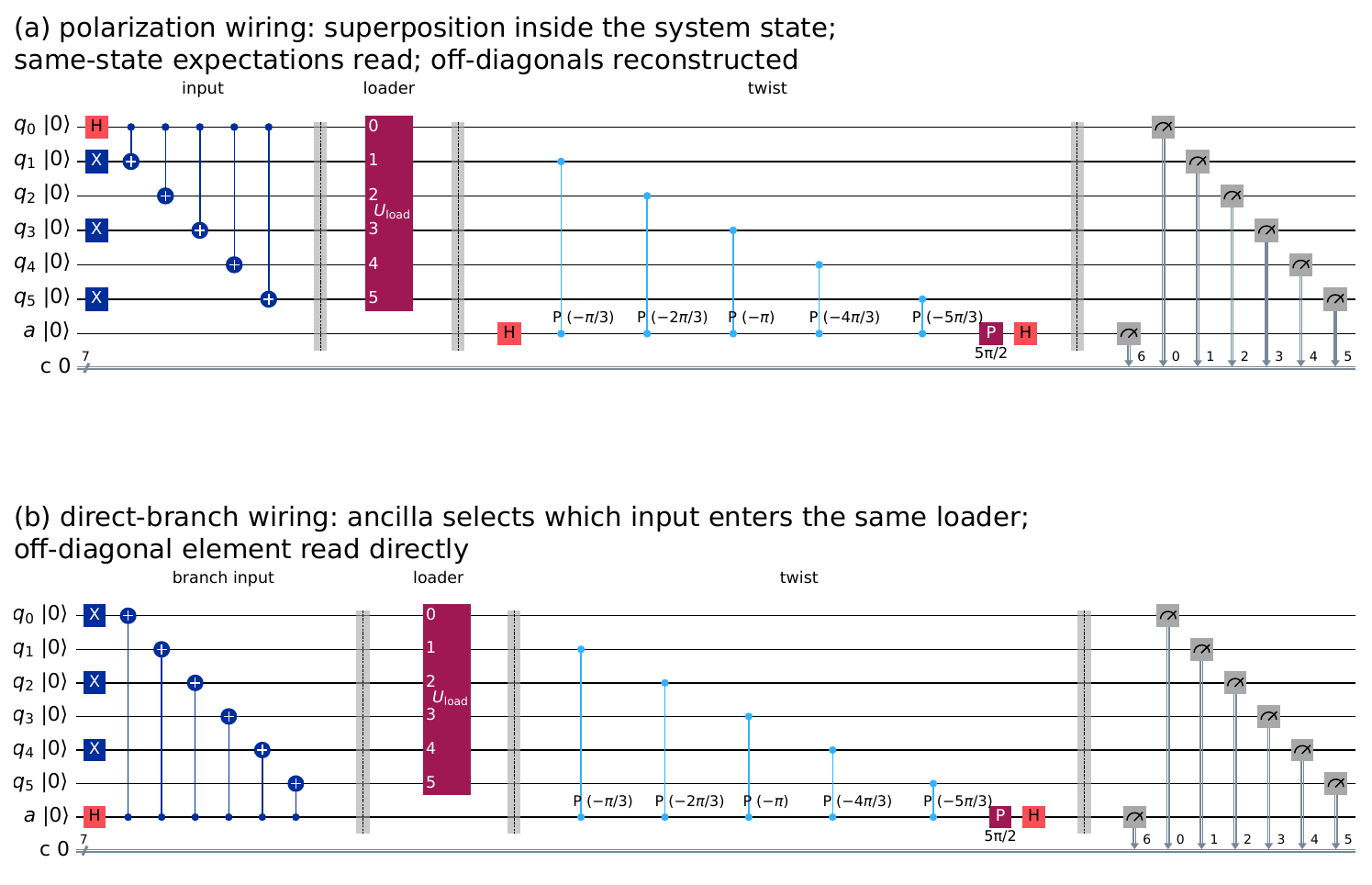}
\caption{\textbf{The two equivalent single-register wirings} ($L=6$,
the setting measuring the real part of the expectation value; the
imaginary part is measured by inserting one $S^\dagger$ on the
ancilla before its closing Hadamard).  (a) Polarization wiring, as
executed for the main-text results [the same circuit as main-text Fig.~2(a)].  (b) Direct-branch wiring. The base input is prepared by the $X$ layer, and the ancilla-controlled CNOT column flips it to the other input on the $|1\rangle$ path. The same unconditional loader acts on both branches by linearity, and the ancilla interferometer reads $\langle\psi_b|U_\theta|\psi_a\rangle$.  Both wirings control only the diagonal twist, a product of controlled-phase gates. In both, the $q=0$ controlled phase has angle zero and is omitted, so the product contains $L-1$ gates.
Table~\ref{tab:S_resources} quantifies their near-identical cost
(104 vs 105 CZ).}
\label{fig:S_wirings}
\end{figure*}

Figure~\ref{fig:S_wirings} shows both wirings as executed-style circuits at
$L=6$.  They share every structural element---one unconditional loader,
only the diagonal twist controlled---and differ only in where the branch
superposition is carried.  In the polarization wiring (a) it is carried by the
system state: a Hadamard and a CNOT fan-out superpose the two product inputs, the ancilla stays idle until the shallow twist, and the
interferometer reads same-state expectations from which the off-diagonal
elements are reconstructed.  In the direct-branch wiring (b) the ancilla
receives the Hadamard first and selects which input string enters the loader (one CNOT per differing site). The two correlated states then propagate along the two
ancilla paths, and the closing Hadamard reads the off-diagonal element
$\langle\psi_b|U_\theta|\psi_a\rangle$ directly, with no reconstruction
step.  The polarization wiring keeps the ancilla exposed only during the twist. The direct-branch wiring is entry-addressable, each matrix element $Z_{ab}$ being read by its own circuit, the property exploited at rank 3 in a separate measurement of individual momentum
matrix elements.

For the two-copy determinant construction, define
\begin{equation}
 |\Psi_-\rangle=\frac{|\psi_0\rangle|\psi_1\rangle-
 |\psi_1\rangle|\psi_0\rangle}{\sqrt2}.
\end{equation}
Direct expansion gives
$\langle\Psi_-|U_\theta\otimes U_\theta|\Psi_-\rangle=\det Z$.
The compiled identity agrees with the direct determinant to $3.40\times10^{-16}$. The successful antisymmetric-sector postselection probability is $1/2$ for both the real- and imaginary-part circuits.  It is a valid alternative but
doubles the system register.  A separately compiled controlled complete
loader is reported only as a stress test: specialized controlled-unitary
synthesis could reduce its cost and would be required before it could be
considered a competitive protocol.

Device topology and routing increase the executed $L=6$ circuits beyond
Table~\ref{tab:S_resources}.  The transpiled Hamiltonian-only circuits
contain 158--174 CZ gates and two-qubit depth 71--102. The fidelity-trained set has the same ranges.  These executed ranges and the logical comparison
should not be conflated.

\section{S6. IBM execution and data record}
Every controlled-twist circuit used 8,000 shots (32,000 for the
energy-trained $L=12$ run), optimization-level-3 transpilation, XY4 dynamical decoupling, measurement twirling
enabled, and gate twirling disabled.  Eight jobs of 48 circuits each enter
the final main-text figures: energy-trained loaders at $L=6$ and $8$
(\texttt{ibm\_marrakesh}) and $L=10$ and $12$ (\texttt{ibm\_kingston}), and
fidelity-compiled loaders at $L=6$ (\texttt{ibm\_kingston}), $L=8$
(\texttt{ibm\_marrakesh}), and $L=10$ and $12$ (\texttt{ibm\_kingston}).
Every per-circuit histogram is stored under an explicit filename. The job identifiers are kept with the archive and are available on request.

\section{S7. Independent reconstruction and uncertainty}
Every reported phase was reconstructed independently from the archived
histograms by stand-alone programs that share no code with the analysis
pipeline.

For each circuit histogram, every bootstrap replicate is sampled independently
from the observed multinomial distribution.  The replicate is propagated
through ancilla expectations, $S^z_{\rm tot}=0$ selection (omitted for the raw-count
values of main-text Fig.~2), the polarization identity,
the determinant, and the wrapped phase.  The reported uncertainty is the
sample standard deviation of 2,000 replicates.  These
intervals quantify shot noise conditional on the measured distributions. They do not include calibration drift or other systematic error.  All 48 measured
points are classified correctly.

\begin{figure*}[t]
\centering
\includegraphics[width=0.9\textwidth]{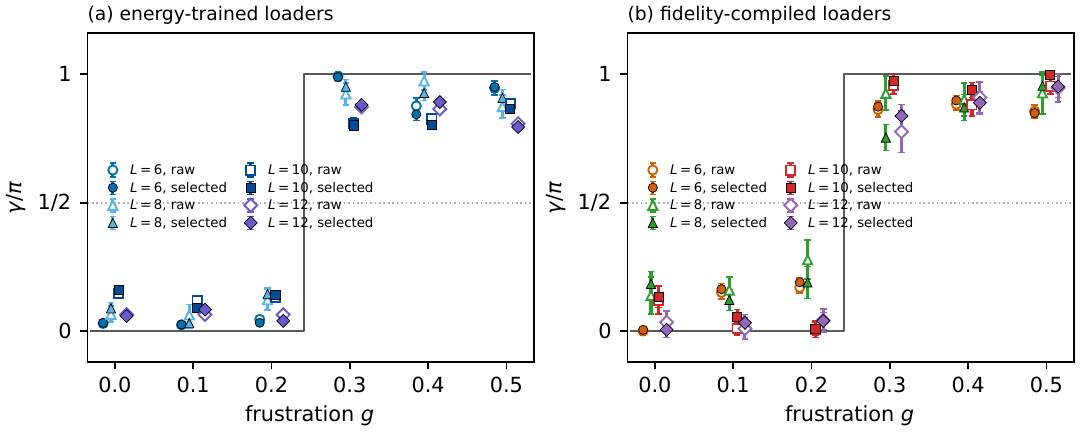}
\caption{Raw (open symbols) and symmetry-selected (filled) reconstructions
of all eight hardware runs: (a) the energy-trained loaders and (b) the
fidelity-compiled loaders, each at $L=6$, $8$, $10$, and $12$.  Points are
horizontally offset for visibility.  All 48 classifications are correct
before and after selection.}
\label{fig:S_robust}
\end{figure*}

Table~\ref{tab:S_robust} and Fig.~\ref{fig:S_robust} compare the reconstruction from all raw shots with the symmetry-selected one for all eight hardware runs. Every run classifies 6/6 both ways, so no classification anywhere in the paper depends on the symmetry filter.  This
tests robustness to that analysis choice, not to calibration drift or
coherent hardware error.
\begin{table*}[t]
\caption{Robustness of the classification to the $S^z_{\rm tot}=0$ filter for all eight
hardware runs: number of correct classifications from all raw shots and
from the postselected shots, range of the retained fraction over the six
couplings, and the smallest distance of $\gamma/\pi$ from the $1/2$
decision boundary.}
\label{tab:S_robust}
\begin{ruledtabular}
\begin{tabular}{llcccc}
run & backend & raw / selected correct & retained fraction & min.\ margin (raw) & min.\ margin (selected)\\
\hline
$L=6$, fidelity & \texttt{ibm\_kingston} & 6/6 / 6/6 & 0.6621--0.6805 & 0.3299 & 0.3095\\
$L=6$, energy & \texttt{ibm\_marrakesh} & 6/6 / 6/6 & 0.6205--0.6494 & 0.3764 & 0.3437\\
$L=8$, fidelity & \texttt{ibm\_marrakesh} & 6/6 / 6/6 & 0.4476--0.4647 & 0.2214 & 0.2532\\
$L=8$, energy & \texttt{ibm\_marrakesh} & 6/6 / 6/6 & 0.3922--0.4136 & 0.3741 & 0.3555\\
$L=10$, fidelity & \texttt{ibm\_kingston} & 6/6 / 6/6 & 0.4840--0.4998 & 0.3791 & 0.3654\\
$L=10$, energy & \texttt{ibm\_kingston} & 6/6 / 6/6 & 0.4536--0.4782 & 0.3091 & 0.3006\\
$L=12$, fidelity & \texttt{ibm\_kingston} & 6/6 / 6/6 & 0.4145--0.4480 & 0.2761 & 0.3373\\
$L=12$, energy & \texttt{ibm\_kingston} & 6/6 / 6/6 & 0.4399--0.4675 & 0.3071 & 0.2943\\
\end{tabular}
\end{ruledtabular}
\end{table*}

Comparing the two complete $L=6$ runs, the Hamiltonian-only postselected values are numerically closer to zero at $g=0.1$ and $0.2$, for example $0.033(11)$ at $g=0.2$ against $0.190(17)$ for the fidelity-trained run, where the parenthesis gives one bootstrap standard deviation in the last digits. At $g=0$ the fidelity-trained value is the closer one.  The two runs were executed on different processors
with different calibrations, so this observation is not interpreted as a
noise-performance advantage of one training objective over the other.  Both
runs independently resolve the change between $g=0.2$ and $g=0.3$.

\section{S8. Diagonal-sampling control}
Because $U_\theta$ is diagonal, each expectation also satisfies $d_k=\sum_zP_k(z)u(z)$, with $P_k(z)$ the probability of the system bit string $z$ in the state $|\chi_k\rangle$ and $u(z)$ the eigenvalue of Eq.~\eqref{eq:S_eigenvalue}.  A separate set of runs therefore omitted the ancilla and
controlled twist, measured the same four coherent system preparations in the computational basis, and evaluated the sum from the recorded counts.  Each circuit used
16,384 shots on \texttt{ibm\_kingston}.  The result is not the main measurement. It is a control on the information content of the readout.

\begin{table*}[h]
\caption{Ancilla-free diagonal-sampling control, reported as postselected
$\gamma/\pi$.  All 24 classifications match the exact reference.}
\label{tab:S_sampling}
\begin{ruledtabular}
\begin{tabular}{ccccccc}
$L$ & $g=0.0$ & $0.1$ & $0.2$ & $0.3$ & $0.4$ & $0.5$\\
\hline
6  & 0.015 & 0.014 & 0.015 & 0.999 & 0.983 & 0.980\\
8  & 0.019 & 0.027 & 0.008 & 0.983 & 0.996 & 0.995\\
10 & 0.008 & 0.002 & 0.029 & 0.987 & 0.970 & 0.996\\
12 & 0.014 & 0.029 & 0.031 & 0.975 & 0.968 & 0.963\\
\end{tabular}
\end{ruledtabular}
\end{table*}

The records of the four sampling jobs retain the raw counts, and their identifiers are kept with the archive.  This control establishes the exact boundary of the claim: coherence
is required in the prepared superposition states and interference was used by
the main circuits, but the diagonal observable is also a functional of
fixed-basis samples.  No quantum advantage or information-theoretic necessity
of the ancilla is inferred.

\section{S9. Single-point label versus loop-based multiplet determinant}
This section backs the main-text statement that, wherever the
loop-based multiplet determinant can be computed for comparison, the
single-point label returns the same quantized value.  The two objects are
different constructions.  The single-point label is the
paper's invariant: the two lowest eigenvectors of the untwisted
Hamiltonian at one coupling, the projected matrix $Z$ of the diagonal ramp
twist $U_\theta$ of Eq.~\eqref{eq:S_twist}, and $\gamma=|\arg\det Z|$.  The loop-based (transported-multiplet, Wilczek--Zee-type) determinant instead gauges the boundary $J_1$ bond $(L{-}1,0)$, whose flip-flop term acquires $e^{\pm i\phi}$. The twist angle $\phi$ is swept around the closed loop $0\to2\pi$ on a uniform grid of $48$ points, and the two lowest eigenvectors are computed at every $\phi$. Consecutive $2\times2$ overlap matrices are polar-projected onto their nearest unitary through the singular-value decomposition, and the ordered product of these unitaries around the loop gives a $2\times2$ holonomy whose determinant phase is the loop label.  The loop label is well defined while the gap between the
two-state subspace and the third level stays open along the entire loop. That gap is recorded at every sampled $\phi$.

Both labels were evaluated by exact diagonalization at every
size--coupling point of the hardware runs, $L=6,8,10,12$ and
$g=0.0,\dots,0.5$.  The result is the same classification at all $24$
points, with both labels exactly quantized:

\begin{table*}[h]
\caption{Single-point label and loop-based multiplet determinant,
$\gamma/\pi$, by exact diagonalization.  At every entry the two independently constructed labels coincide. The single-point label is pinned to $\{0,\pi\}$ within $10^{-15}$ and the loop label within
$2\times10^{-14}$ of the same value.}
\label{tab:S9}
\begin{ruledtabular}
\begin{tabular}{ccccccc}
$L$ & $g=0.0$ & $0.1$ & $0.2$ & $0.3$ & $0.4$ & $0.5$\\
\hline
6  & 0 & 0 & 0 & 1 & 1 & 1\\
8  & 0 & 0 & 0 & 1 & 1 & 1\\
10 & 0 & 0 & 0 & 1 & 1 & 1\\
12 & 0 & 0 & 0 & 1 & 1 & 1\\
\end{tabular}
\end{ruledtabular}
\end{table*}

Three checks accompany the table.  First, validity: the minimum isolation gap above the two-state subspace on the sampled twist grids is $0.047$, attained at $L=12$ and $g=0.2$. No gap closing was observed at any sampled angle. The two levels within the doublet do cross and split along the loop, with a splitting up to $0.68$. This obstructs any single-state loop phase but not the multiplet product.  Second, discretization: halving
the grid from $48$ to $24$ points changes no loop label by more than
$2\times10^{-14}$.  Third, magnitude: the single-point determinant
magnitudes $|\det Z|$ lie between $0.42$ and $0.66$ across all $24$
points, far from the zero at which a phase would degrade.  The agreement
is a physics statement rather than an algebraic identity: a global ramp
evaluated at one Hamiltonian point and a transported family over a
boundary-bond loop are different constructions, and both detect the same
quantized classification of the transition.  The cost asymmetry is the
point of the comparison: the loop label consumed $48$ diagonalizations
per size--coupling point where the single-point label consumed one.